\documentclass[a4paper,11pt]{article}
\usepackage{pos}

\usepackage[normalem]{ulem}
\usepackage{color}

\title{
\vspace{-5ex}     
\begin{flushright} {\normalsize LFTC-21-5/66} 
\end{flushright}
\vspace{5ex}
\boldmath{$\Upsilon$} and \boldmath{$\eta_b$} mass shifts in nuclear matter
and the $^{12}${\rm C} nucleus bound states}

\author*[a]{G.~N.~Zeminiani}
\author[b]{J.J. Cobos-Mart\'inez}
\author[a]{K.~Tsushima}

\affiliation[a]{Laborat\'orio de F\'isica Te\'orica e Computacional, 
Universidade Cidade de S\~ao Paulo (UNICID), and Universidade Cruzeiro do Sul,\\
01506-000, S\~ao Paulo, SP, Brazil}

\affiliation[b]{Departamento de F\'isica, Universidad de Sonora,\\
Boulevard Luis Encinas J. y Rosales, Colonia Centro, Hermosillo, Sonora
83000, M\'exico}

\emailAdd{guilherme.zeminiani@gmail.com}
\emailAdd{jesus.cobos@fisica.uson.mx}
\emailAdd{kazuo.tsushima@cruzirodosul.edu.br}

\abstract{This is a contribution for the PANIC 2021 Proceedings 
based on the articles, Eur. Phys. J. A 57, 259 (2021) and  
the accompanied article $[$arXiv:2109.08636 $[$hep-ph$]]$ (Hadron 2021 contribution). 
We have estimated for the first time the mass shifts of the $\Upsilon$ and $\eta_b$ mesons in 
symmetric nuclear matter by an SU(5) flavor symmetric effective Lagrangian approach, as well as the 
in-medium mass of $B^*$ meson by the quark-meson coupling (QMC) model.
The attractive potentials for the $\Upsilon$- and $\eta_b$-nuclear matter are obtained, 
and one can expect for these mesons to form nuclear bound states. 
We have indeed found such nuclear bound states with $^{12}$C nucleus, 
where the results for the $^{12}$C nucleus bound state energies are new, 
and we report here for the first time.
}

\FullConference{%
  *** Particles and Nuclei International Conference - PANIC2021 ***\\
  *** 5 - 10 September, 2021 ***\\
  *** Online ***
}

\begin{document}
\maketitle

\section{Introduction}
By studying the $\Upsilon$- and $\eta_b$-nucleus interactions, 
one can advance in understanding the heavy meson and heavy quark interactions with nucleus 
based on quantum chromodynamics (QCD). For a possible mechanism of the 
bottomonium interaction with the nuclear medium (nucleus), we apply here, via the 
excitations of the intermediate state hadrons which contain the light quarks $u$ and $d$.

We first calculate the in-medium $B$ and $B^{*}$ meson masses, then we estimate 
the mass shifts of the $\Upsilon$ and $\eta_b$ mesons through the 
excitations of intermediate state $B$ and $B^*$ mesons in the $\Upsilon$ and $\eta_b$ 
self-energies. 
The estimates will be made by an SU(5) effective Lagrangian which contains both the 
$\Upsilon$ and $\eta_b$ mesons with one universal coupling constant.
Thus, we need to know better the $B$ and $B^{*}$ meson properties in medium. 
For this purpose we use the quark-meson coupling (QMC) model invented by 
Guichon~\cite{Guichon:1987jp}, which has been successfully applied for various 
studies~\cite{Saito:2005rv,Krein:2017usp}.

The interesting question is, whether or not the attractive $\Upsilon$- 
and $\eta_b$-nuclear matter 
interactions are strong enough to form nuclear bound states. 
Thus, we study the $\Upsilon$- and $\eta_b$-$^{12}$C bound states for the first time.

\section{$\Upsilon$ and $\eta_b$ mass shifts in symmetric nuclear matter}
We calculated the effective masses (Lorentz scalar) of the $B$ and $B^{*}$ mesons in symmetric 
nuclear matter using the QMC model, where that of the  
$B^{*}$ meson was the first time~\cite{Zeminiani:2020aho}.

\begin{figure}[htb]
\vspace{0.5cm}
\centering
 \includegraphics[width=6.4cm]{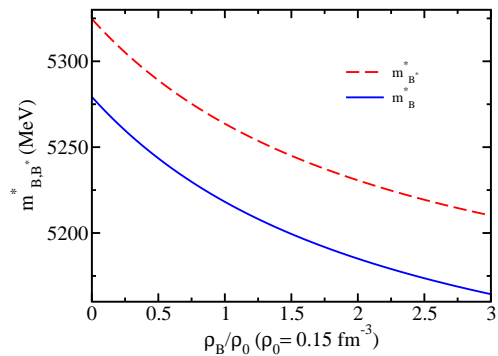}
 \caption{$B$ and $B^{*}$ meson effective masses (Lorentz scalar)  
 in symmetric nuclear matter. 
}
 \label{fig1}
\end{figure}

As shown in Fig.~\ref{fig1}, the QMC model predicts a similar amount of the $B$ and $B^*$ mass 
shifts. The mass shifts predicted are, respectively, 
$(m^*_B - m_B)=-61$ MeV and $(m^*_{B^*}-m_{B^*})=-61$ MeV at $\rho_0 = 0.15$ fm$^{-3}$, 
with the difference in the next digit.
To calculate the $\Upsilon$ and $\eta_b$ meson self-energies
in symmetric nuclear matter via the $B$ and $B^{*}$ meson
loops, we use the obtained in-medium masses shown in Fig.~\ref{fig1}.

The $\Upsilon$ and $\eta_b$ mass shifts in medium come from the modifications of the $BB$, $BB^{*}$ 
and $B^{*}B^{*}$ meson loop contributions to their self-energies, relative to those in free space,
where the self-energies are calculated based on an
effective flavor SU(5) symmetric Lagrangian, with the one SU(5) universal coupling constant value 
determined by the vector meson dominance (VMD) hypothesis (model) with the experimental data for
$\Gamma (\Upsilon \rightarrow e^{+}e^{-})$~\cite{Zeminiani:2020aho}.
We use phenomenological form factors
to regularize the self-energy integrals,
which are dependent on the cutoff $\Lambda_{B} = \Lambda_{B^{*}}$ values   
in the range $2000~\text{MeV} \leq \Lambda_{B,B^*} \leq 6000~\text{MeV}$.

For our predictions, we take the minimum meson loop contributions,
namely, that is estimated by including only the $BB$ meson loop for the $\Upsilon$ self-energy, 
and only the $BB^{*}$ meson loop for the $\eta_b$ self-energy. 
This is necessary, because the unexpectedly larger contribution from 
the heavier $B^*B^*$ meson loop was observed~\cite{Zeminiani:2020aho}.
Note that, we ignore the possible widths, or the imaginary parts in the self-energies in the present study.
We plan, however, to include the effects of the widths into the calculation in the near future.

\begin{figure}[!htb]
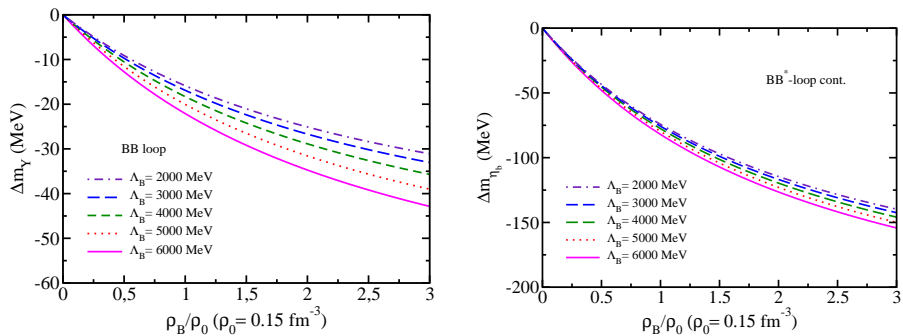
%
\centering
\includegraphics[width=5.6cm]{1_BB.eps}
\hspace{2ex}
\includegraphics[width=5.6cm]{etab_BBs.eps}
\caption{$BB$ loop contribution to the 
$\Upsilon$ mass shift (left) and $BB^{*}$ loop contribution to the 
$\eta_b$ mass shift (right) for five different values of the cutoff 
mass $\Lambda_{B} (=\Lambda_{B^*})$.}
\label{fig2}
\end{figure}

The calculated mass shifts of the $\Upsilon$ and $\eta_b$ mesons are shown in 
Fig.~\ref{fig2}.
As one can see in the left panel for the $\Upsilon$, the effect of the decrease in the $B$ meson 
in-medium mass yields a negative mass shift of the $\Upsilon$. 
The decrease of the $B$ meson mass in nuclear
matter enhances the $BB$ meson loop contribution for the $\Upsilon$ in-medium self-energy 
in such a way to yield a negative mass shift, which is also dependent on the 
cutoff mass value $\Lambda_B$. Namely, the amount of the mass shift increases as $\Lambda_B$ 
increases, ranging from -16 to -22 MeV at symmetric nuclear matter 
saturation density, $\rho_0$.
For the $\eta_b$ mass shift, which is estimated  
by including only the $BB^*$ meson loop (right panel), it 
ranges from -75 to -82 MeV at $\rho_0$ for the five cutoff mass values, the same  
as those for the $\Upsilon$.

As one can see, the mass shift of $\eta_b$ is larger than that of the $\Upsilon$.
This reflects the fact that the $\eta_b$ interaction Lagrangian   
has a larger number of interaction terms contributing to the self-energy,  
and results to yield more contribution than that of the $\Upsilon$. 
The use of the SU(5) symmetric couplings  
also gives an impact on the calculated $\eta_b$ mass shift, as well as   
on the $^{12}$C nucleus bound states energies to be given in the next section.

\vspace{0.5cm}

\section{$\Upsilon$- and $\eta_b$-nucleus bound states with $^{12}$C}
\begin{figure}[htb]
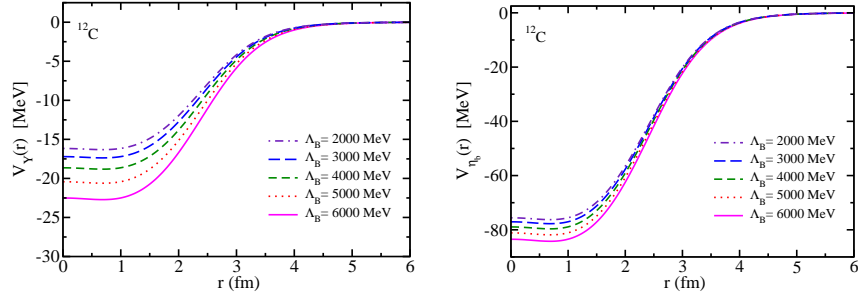
%
\centering
\includegraphics[width=5.3cm]{Upsipot_C12.eps}
\hspace{2ex}
\includegraphics[width=5.3cm]{etabpot_C12.eps}
\caption{$\Upsilon$- and $\eta_b$-nucleus potential for the
$^{12}$C nucleus.}
\label{fig3}
\end{figure}

We consider the situation that an $\Upsilon$ or an $\eta_b$ meson is produced inside
a $^{12}$C nucleus with nearly zero relative momentum to $^{12}$C, where the $^{12}$C has baryon 
density distribution $\rho^{^{12}\text{C}}_{B}(r)$, and we follow the procedure of  
Refs.~\cite{Tsushima:2011kh,Cobos-Martinez:2020ynh}.
In Ref.~\cite{Zeminiani:2021xvw} we have presented the result for the $^4$He case, where the 
density profile was parameterized and taken from~\cite{Saito:1997ae}.
However, for the $^{12}$C nucleus in the present case, the density profile is calculated by the 
QMC model. We also use a local density approximation to obtain the $\Upsilon$ and $\eta_b$ nuclear 
potentials inside the $^{12}$C nucleus, which are shown in Fig.~\ref{fig3} 
for five values of $\Lambda_B$. The potentials are both attractive, 
with their depths depending on the  
cutoff mass values, namely, the deeper the larger $\Lambda_B$.

\vspace{0.5cm}

The $\Upsilon$- and $\eta_b$-$^{12}$C bound state energies are then
calculated by solving the Klein-Gordon equation using the nuclear potentials 
shown in Fig.~\ref{fig3}. Although $\Upsilon$ is a spin-1 particle, we make an approximation that  
the transverse and longitudinal components in the Proca equation 
are expected to be very similar for the $\Upsilon$ nearly at rest, hence it 
is reduced to one component, which corresponds to the Klein-Gordon equation.
The bound state energies are calculated for the same values of the cutoff mass $\Lambda_B$ used
in the previous section, and the results are given in Table.~\ref{tab:upsilon-C12-kge}.
Note that, due to the large number of the $\eta_b$ bound states found for $^{12}$C,
we have not shown the shallower bound state energies explicitly for the $\eta_b$ in the table. 
The results indicate that both the $\Upsilon$ and $\eta_b$ are expected to form bound states with 
the $^{12}$C nucleus.
We will consider other nuclei in the upcoming study~\cite{Zeminiani:2022amo}.
We emphasize that, even though the values of the bound state energies vary according 
to the chosen values of the cutoff mass, the prediction that the $\Upsilon$ and $\eta_b$ 
are expected to form bound states with the $^{12}$C nucleus, is independent of the values chosen.
By ignoring the widths, the experimental observation of the predicted
bound states could be an issue, but the present study
primarily focuses on the existence of the bound states.
We plan to include the effects of the widths
in the future study~\cite{Zeminiani:2022amo} to see the impact of them on the results.

\begin{table}[htb]
  \caption{\label{tab:upsilon-C12-kge} $^{12}_{\Upsilon}\text{C}$ and $^{12}_{\eta_b}\text{C}$ bound state
  energies. When $|E| < 10^{-1}$ MeV we consider there is no bound state,
  which we denote with ``n''. All dimensioned quantities are in MeV.
  The shallower bond states for the $\eta_b$ are not shown explicitly.
}
\begin{center}
\scalebox{0.7}{
\begin{tabular}{ll|r|r|r|r|r}
  \hline \hline
  & & \multicolumn{5}{c}{Bound state energies} \\
  \hline
& $n\ell$ & $\Lambda_{B}=2000$ & $\Lambda_{B}=3000$ & $\Lambda_{B}= 4000$ &
$\Lambda_{B}= 5000$ & $\Lambda_{B}= 6000$ \\
\hline
$^{12}_{\Upsilon}\text{C}$
& 1s & -10.6 & -11.6 & -12.8 & -14.4 & -16.3 \\
& 1p & -6.1 & -6.8 & -7.9 & -9.3 & -10.9 \\
& 1d & -1.5 & -2.1 & -2.9 & -4.0 & -5.4 \\
& 2s & -1.6 & -2.1 & -2.8 & -3.8 & -5.1 \\
& 2p & n & n & n & -0.1 & -0.7 \\
\hline
$^{12}_{\eta_b}\text{C}$
& 1s & -65.8 & -67.2 & -69.0 & -71.1 & -73.4 \\
& 1p & -57.0 & -58.4 & -60.1 & -62.1 & -64.3 \\
& 1d & -47.5 & -48.8 & -50.4 & -52.3 & -54.4 \\
& ...& ... & ... & ...& ...& ... \\
& 2h & n & n & n & -0.2 & -1.2 \\
\hline
\end{tabular}
}
\end{center}
\end{table}

\section{Summary and Conclusion}
We have estimated for the first time the $B^*$, $\Upsilon$ and $\eta_b$ mass shifts 
in symmetric nuclear matter, as well as the $\Upsilon$- and $\eta_b$-$^{12}$C 
bound state energies neglecting any possible widths of the mesons,  
assuming each meson is produced inside the $^{12}$C nucleus with nearly zero relative 
momentum.

The in-medium $B$ and $B^*$ meson masses necessary to evaluate the 
$\Upsilon$ and $\eta_b$ self-energies are calculated by 
the quark-meson coupling model.
Our predictions, taking only the $BB$ meson loop contribution for the $\Upsilon$ mass shift,
and only the $BB^{*}$ meson loop contribution for the $\eta_b$ mass shift, 
give the $\Upsilon$ mass shift that varies from -16 MeV to -22 MeV at symmetric nuclear matter 
saturation density ($\rho_0 = 0.15$ fm$^{-3}$) for the cutoff mass values in the range from 2000 MeV 
to 6000 MeV, while for the $\eta_b$ it ranges from -75 to -82 MeV at $\rho_0$ 
for the same cutoff mass value range.  

For the $\eta_b B B^*$ coupling constant value, we have used the SU(5) 
universal coupling constant and the value determined by the $\Upsilon BB$ coupling constant   
by the vector meson dominance model with the experimental data.

For the $\Upsilon$ or $\eta_b$ meson produced inside the $^{12}$C nucleus with nearly zero 
relative momentum to $^{12}$C, their attractive interactions are strong enough to form bound 
states with the $^{12}$C nucleus. 
The bound state energies have been calculated by solving
the Klein-Gordon equation, with the nuclear potentials obtained using a local density approximation, 
and the nuclear density distribution calculated by the quark-meson coupling model.

We plan to elaborate the present study in the near future by including the effects of the widths,
as well as using different regularization methods and/or form factors in the 
$\Upsilon$ and $\eta_b$ self-energies.

\vspace{2ex}
\noindent
{\bf Acknowledgements:}\\
This work was supported by CAPES-Brazil (GNZ), 
CNPq-Brazil, Process, No.~313063/2018-4, process, No.~426150/2018-0, 
and FAPESP-Brazil, Process, No.~2019/00763-0 (KT), 
and was also part of the projects INCT-FNA-Brazil, 
Process, No.~464898/2014-5.



\end{document}